# Squeezing electromagnetic energy with a dielectric split ring inside a permeability-near-zero metamaterial

Yi Jin[1], Pu Zhang[1,2] and Sailing He[1,2,*]

[1]*Centre for Optical and Electromagnetic Research, State Key Laboratory of Modern Optical Instrumentations, Joint Research Center of Photonics of the Royal Institute of Technology (Sweden) and Zhejiang University, Zijingang Campus, Hangzhou 310058, China*

[2]*Division of Electromagnetic Engineering, School of Electrical Engineering, Royal Institute of Technology, S-100 44 Stockholm, Sweden*

A novel electromagnetic energy squeezing mechanism is proposed based on the special properties of permeability-near-zero metamaterials. It is found that nearly no energy stream can enter a simply-connected conventional dielectric material positioned inside a permeability-near-zero material. When the dielectric domain is shaped as a split ring (with a gap opened) surrounding a source, the electromagnetic energy generated by the source is forced to propagate through the gap. When the gap is narrow, the energy stream density becomes very large and makes the magnetic field enhanced drastically in the gap. The narrow gap can be long and bended. This provides us a method to obtain strong magnetic field without using resonance enhancement.

Based on the theoretical work of Pendry *et al*. [1,2], it was experimentally demonstrated at microwave frequencies by Smith *et al*. in 2000 [3] that an artificial microstructured composite (based on arrays of metallic split ring resonators and wires) could exhibit a negative refractive index. This demonstration has attracted many scientists to explore these artificial materials which are usually termed as metamaterials. More and more exotic phenomena and applications, not easily achievable using naturally occurring materials, have been disclosed and justified [4−7], as metamaterials can provide unusual values (e.g., negative, very small or very large) for permittivity and permeability. Usually, the permittivity and permeability of a metamaterial follow the Drude or Lorentz dispersion model. Near the plasma frequency, their values can be near zero or even zero. The research topic of metamaterials with low permittivity and/or permeability has also attracted much attention recently [8−14]. In the early stage, such a metamaterial was mainly used to get high radiation directivity [9]. Lately, Engheta *et al*. suggested several applications of near zero permittivity, such as shaping the phase front [10] and transmitting subwavelength image [11]. They have also shown that the electromagnetic wave in a wide metallic waveguide can be tunneled through a very narrow metallic waveguide filled with a permittivity-near-zero material [12-14], and strong electric field is obtained. To enhance magnetic field, one may use the dual configuration of what Engheta *et al*. have proposed, however, a perfect magnetic conductor will then be required as the boundary. In this paper, we will show a novel energy squeezing mechanism to enhance magnetic field. The squeezing system is open without using a perfect magnetic conductor as the boundary, in which only permeability-near-zero and conventional dielectric materials are used. The key point of this squeezing mechanism is that nearly no energy stream can enter a simply-connected conventional dielectric material positioned inside a permeability-near-zero material. In the following, the new squeezing method will be explained.

Consider two-dimensional (2D) electromagnetic propagation with the electric field perpendicular to the *x-y* plane (TE polarization). Maxwell's equations are simplified as

$$\frac{\partial E_z(x,y)}{\partial y}\mathbf{e}_x - \frac{\partial E_z(x,y)}{\partial x}\mathbf{e}_y = i\omega\mu(H_x(x,y)\mathbf{e}_x + H_y(x,y)\mathbf{e}_y), \quad (1.a)$$

$$\frac{\partial H_y(x,y)}{\partial x} - \frac{\partial H_x(x,y)}{\partial y} = -i\omega\varepsilon E_z(x,y), \quad (1.b)$$

where time harmonic factor *exp(−iωt)* has been removed from the electromagnetic field quantities. When $\mu$ tends to zero, one has

$$\frac{\partial E_z(x,y)}{\partial x} = \frac{\partial E_z(x,y)}{\partial y} = 0. \quad (1)$$

This means that the electric field is uniform in the whole $\mu=0$ domain. It should be noted that the magnetic field in the domain of $\mu=0$ is usually not uniform, and its phase may vary at different positions.

Now consider the configuration shown in Fig. 1. Domain 1 of zero permeability $\mu_1$ is surrounded by exterior domain 4 of free space. Domains 2 and 3 of conventional dielectric material (free space here) are positioned inside domain 1. The shapes of domains 1−3 can be arbitrary. There is a current source propagating along the $z$ axis in domain 2, and no source in domain 3. Let $F_n(x,y)$ denote some quantity $F(x,y)$ in domain $n$ ($n$=1, 2, 3, 4). Electric field $E_{1,z}(x,y)$ in domain 1 is uniform, and thus assumed to be constant $E_z$. The value of $E_z$ can be obtained as follows. In domain 1, we can rewrite Eq. (1) in the following integration form

$$\oint_{\partial 1} \mathbf{H}_{1,\partial 1}(x,y)\cdot d\mathbf{l} + \oint_{\partial 2} \mathbf{H}_{1,\partial 2}(x,y)\cdot d\mathbf{l} + \oint_{\partial 3} \mathbf{H}_{1,\partial 3}(x,y)\cdot d\mathbf{l} = -i\omega\varepsilon_0 \int_1 E_{1,z}(x,y)ds, \quad (2)$$

where integration contours $\partial 1$, $\partial 2$ and $\partial 3$ are shown in Fig. 1, $\mathbf{H}_{1,\partial n}(x,y)$ represents the value of magnetic field $\mathbf{H}_1(x,y)$ at boundary $\partial n$, and $\varepsilon_0$ is the vacuum permittivity. The right side of Eq. (3) can be written as $-iA_1\omega\varepsilon_0 E_z$, where $A_1$ is the area of domain 1. According to the continuity condition of the tangential electric and magnetic fields at both sides of a boundary [15], electric fields $E_{2,z,\partial 2}(x,y)$, $E_{3,z,\partial 3}(x,y)$, and $E_{4,z,\partial 1}(x,y)$ are all equal to $E_z$, and the contour integrals of $\mathbf{H}_{1,\partial n}(x,y)$, $n$=1, 2, 3, at the left side of Eq. (3) can be replaced by those of $\mathbf{H}_{4,\partial 1}(x,y)$, $\mathbf{H}_{2,\partial 2}(x,y)$, and $\mathbf{H}_{3,\partial 3}(x,y)$, respectively. Following the unique theorem [15], if one knows the tangential electric field at the boundary of a domain (note that the tangential magnetic field is not required) and the excitation source in the domain, the electric and magnetic fields in the whole domain are uniquely determined. Thus, the three terms at the left side of Eq. (3) can be formally written as $f_4(E_z)$, $f_2(E_z)$, and $f_3(E_z)$. The current source in domain 2 is taken into account in $f_2(E_z)$. Based on the above discussion, Eq. (3) can be written as follows,

$$f_2(E_z) + f_3(E_z) + f_4(E_z) = -iA_1\omega\varepsilon_0 E_z. \quad (3)$$

It is noted that the expressions of $f_2(E_z)$, $f_3(E_z)$ and $f_4(E_z)$ do not depend on the positions of domains 2 and 3 inside domain 1. That is, with a precondition that domains 1−3 do not cross mutually, the positions of domains 2 (with the same relative position of the current source) and 3 do not influence the expressions of $f_2(E_z)$, $f_3(E_z)$, and $f_4(E_z)$. Thus, the value of $E_z$ obtained from Eq. (4) does not depend on the positions of domains 2 and 3. Consequently, the electromagnetic fields in domains 2−4 do not depend on the positions of domains 2 and 3, either. This is a very interesting result brought by zero permeability. When we move domain 2 and/or domain 3 inside domain 1, however, the magnetic field in domain 1 will change, even drastically, because the structure of domain 1 changes. The above results can be extended to a more general case when there are several arbitrary domains of conventional materials positioned inside domain 1, some of which may possess internal current sources.

Since permeability $\mu_1$ is zero, magnetic induction $\mathbf{B}_1(x,y)$ [equal to $\mu_1 \mathbf{H}_1(x,y)$] in domain 1 is zero. Following the continuity condition of the normal magnetic inductions at both sides of boundary $\partial 3$ [15], one knows that the normal component of $\mathbf{B}_{3,\partial 3}(x,y)$ and consequently that of $\mathbf{H}_{3,\partial 3}(x,y)$ in domain 3 are zero at boundary $\partial 3$. Thus, Poynting vector $\mathbf{P}_{3,\partial 3}(x,y)$ is normal to boundary $\partial 3$. However, its time average, $\mathbf{S}_{3,\partial 3}(x,y)$, is zero. In fact, energy stream density $\mathbf{S}_3(x,y)$ is zero in the whole domain 3. This is shown as follows. Electric field $E_{3,z}(x,y)$ in domain 3 satisfies the following Helmholtz equation,

$$\Delta E_{3,z}(x,y) + \mu_0 \varepsilon_0 \omega^2 E_{3,z}(x,y) = 0, \tag{5}$$

where $\mu_0$ is the vacuum permeability. At boundary $\partial 3$, $E_{3,z}(x,y)=E_z$ is constant. $E_{3,z}(x,y)$ can be written as $E_{3,z}(x,y)=real(E_{3,z}(x,y))+i*imag(E_{3,z}(x,y))$. Then Eq. (5) can be rewritten as

$$\Delta real(E_{3,z}(x,y)) + \mu_0\varepsilon_0\omega^2 real(E_{3,z}(x,y)) = 0, \qquad (6.a)$$

$$\Delta imag(E_{3,z}(x,y)) + \mu_0\varepsilon_0\omega^2 imag(E_{3,z}(x,y)) = 0. \qquad (6.b)$$

Eqs. (6.a) and (6.b) require domain 3 to be lossless. From Eqs. (6.a) and (6.b), one sees that the real and imaginary parts of $E_{3,z}(x,y)$ follow two independent equations, and they are correlated only through the boundary condition. First we assume that $E_z$ is equal to 1, then $imag(E_{3,z}(x,y))$ is zero at boundary $\partial 3$. Consequently, it can be deduced that the imaginary part of $E_{3,z}(x,y)$ fulfilling Maxwell's equations should be zero. Otherwise, it can be assumed to be $p(x,y)$ [$p(x,y)$ is real and zero at boundary $\partial 3$]. $p(x,y)$ multiplied with an arbitrary number $c$ can still fulfill Eq. (6.b), and the product is zero at boundary $\partial 3$. Then, $E_{3,z}(x,y)+c*p(x,y)$ is also a solution of Maxwell's equations, which is still equal to $E_z$ at boundary $\partial 3$. This is not consistent with the unique theorem [15], according to which, there is only a unique solution in domain 3 when the tangential electric field at boundary $\partial 3$ is given. Thus, $imag(E_{3,z}(x,y))$ should be zero. Since $E_{3,z}(x,y)$ is real, it can be written as $E_{3,z}(x,y)=q(x,y)$ [$q(x,y)$ is real]. When $E_z$ is not equal to 1, it is easily deduced that $E_{3,z}(x,y)= E_z q(x,y)$. With Eq. (1), one has

$$H_{3,x}(x,y) = -\frac{iE_z}{\omega\mu_0}\frac{\partial q(x,y)}{\partial y}, \qquad (7.a)$$

$$H_{3,y}(x,y) = \frac{iE_z}{\omega\mu_0}\frac{\partial q(x,y)}{\partial x}. \qquad (7.b)$$

When $E_{3,z}(x,y)$, $H_{3,x}(x,y)$, and $H_{3,y}(x,y)$ are put into the expression of energy stream $\mathbf{S}_3(x,y)$, one has

$$S_{3,x}(x,y) = \frac{1}{2} real(-E_{3,z}(x,y)H_{3,y}(x,y)^*) = 0, \tag{8.a}$$

$$S_{3,y}(x,y) = \frac{1}{2} real(E_{3,z}(x,y)H_{3,x}(x,y)^*) = 0. \tag{8.b}$$

Based on the above discussion, one sees that energy stream $\mathbf{S}_3$ is zero in domain 3. There is no energy stream (averaged in time) flowing through any point of boundary ∂3. This is the most interesting result, which is different from a usual case in which some electromagnetic energy stream enters a domain through some part of the boundary, and then goes out of the domain through another part of the boundary.

As shown above, no electromagnetic energy can enter a conventional lossless dielectric material inside a material of zero permeability. Consequently, if this domain with a gap opened (like domain 3 of a broken ring shape shown in Fig. 1) encloses a source region, the generated electromagnetic energy can go nowhere else but propagate through the gap. The case of a narrow gap is interesting. If the power generated by the source can be maintained at a similar level while narrowing the gap gradually, the energy stream density in the gap will increase to a very large value in the end, as shown below. Slight reduction of the width of the gap will change little the whole shape of domain 3. Thus, $f_3(E_z)$ in Eq. (4) varies little. The slight variation of $f_3(E_z)$ with $f_2(E_z)$ and $f_4(E_z)$ fixed usually causes only small change to the value of $E_z$ obtained from Eq. (4), and the electromagnetic fields in domains 2−4 (which depend on $E_z$). Then, the power emitted into the exterior space changes little, and the power generated by the current source, which is equal to the former, also changes little. On the other hand, the slight reduction of the width can be *relatively* large with respect to the width of a narrow gap. This can cause the energy stream density increase quickly in the narrow gap. Because the electric field in the gap (equal to $E_z$) changes little, the magnetic field there must be enhanced greatly to increase the energy stream density. Both the energy stream density and the magnetic field amplitude are nearly inversely proportional to the width of the narrow gap.

To illustrate and verify the proposed energy squeezing mechanism, numerical simulation is carried out with a finite-element-method software, Comsol [16]. The configuration of Fig. 1 is used. The wavelength in free space is $0.3a$ ($a$ is the length unit). The widths of domains 1 and 2 are $a_1=a$ and $a_2=0.21a$, respectively. The width and thickness of domain 3 are $a_3=0.75a$ and $w=0.125a$, respectively. Domain 2 can be arbitrarily positioned in the region surrounded by dielectric domain 3 of a split ring shape, whose center overlaps with those of domains 1 and 2 in our simulation. The current source is a line current at the center of domain 2. In simulation, the four corners of the gap are filleted to form rounded corners with a radius of $0.01a$ in order to weaken possible strong localized magnetic field around them, and the permeability of domain 1 is set to $\mu_1 = 10^{-5}+10^{-4}i$ in order to show that the squeezing mechanism can still work when $\mu_1$ has a small deviation from zero. Figs. 2(a) and 2(b) show the distributions of normalized electric and magnetic amplitudes, respectively, when the width of the gap is $g=0.06a$. In this paper, when we say that some quantity is normalized, the quantity is normalized by the corresponding one at the center of the gap when domain 1 is replaced by free space (i.e., when the line current is in free space). In Fig. 2(a), one can see that the electric field in domain 1 is rather uniform, as expected. In contrast, as shown in Fig. 2(b), the magnetic field in domain 1 can be quite non-uniform, and is strongly enhanced in the gap. Fig. 2(b) also shows the energy stream density with arrows. Outside but near domain 3, the arrows circulate along boundary $\partial 3$. In domain 3, energy stream density $\mathbf{S}_3(x,y)$ has a small value due to nonzero $\mu_1$, and can be negligible. In the gap, the energy stream density arrows are nearly parallel to the side edges, from which one can see that the $y$ component of the magnetic field is enhanced greatly. As the width of the gap varies, Fig. 2(c) shows the normalized electric and magnetic amplitudes at the center of the gap, and Fig. 2(d) shows the normalized power generated by the line current and the normalized power emitted into the exterior space as well as the normalized energy stream density at the center of the gap. During this process, the energy stream density and magnetic amplitude at the center of the gap are approximately inversely proportional to the width of the gap as expected, whereas the electric amplitude does not change much. When $g=0.06a$, the energy stream density and magnetic amplitude have been enhanced by about 57 and 260 times, respectively. As the gap becomes narrower, the stronger magnetic field makes the material loss larger in the gap, which is compensated by more power generated by the line current, and thus the power emitted into the exterior space is reduced little. Finally, it should be noted that the power generated by the current source may be influenced by the environment, especially domain 2. By adjusting the width of domain 2 appropriately, the absolute value of the energy stream density and magnetic amplitude in the gap can be increased further. This is not discussed here since the main focus of the present paper is the mechanism and process of squeezing energy.

For energy squeezing, the narrow gap of domain 3 can be long and bended. Fig. 3 gives such an example when domain 3 is transformed into a spiral with a long narrow gap (see the solid lines). All the parameters are the same as those used in Fig. 2(a) except the shape of domain 3. The width of the gap is still $0.06a$. As shown in Fig. 3(a), the electric field is a bit non-uniform in domain 1. This is because the strong electromagnetic wave will be attenuated gradually by the material loss when propagating along the long gap. Nevertheless, the energy squeezing mechanism still works well. As shown in Fig. 3(b), the energy from the source is squeezed into the long bended gap and streams along it. The magnetic field is enhanced strongly in the whole gap, although it is weakened gradually due to the material loss.

Finally, we note that if the gap of domain 3 is further narrowed to be closed, the energy squeezing phenomenon will disappear. The topology of domain 1 changes completely, and two individual parts are isolated by domain 3. The electric field is still uniform in each of the two individual parts, but their values may be different in different parts. Eq. (4), which requires the whole region of domain 1 to be connected, is no longer valid for this case. Thus, $\mathbf{S}_{3,\partial 3}(x,y)$ may not be zero at boundary $\partial 3$, and the energy from the current source can go into and through domain 3. However, if we make a narrow enough air slit crossing the narrow gap of domain 3, energy squeezing phenomenon can still be obtained. This is illustrated in Fig. 4 for a case when a 0.002$a$-width air slit is introduced in the middle crossing the gap of the structure in Fig. 2(a). As shown in Fig. 4(b), a large part of the electromagnetic energy from the source can still be squeezed through the narrow gap, and strong magnetic field exists in both the narrow gap and the air slit. Such an air slit can be used conveniently for inserting some material for nonlinearity or sensing application. Some air slit with enhanced field was also used by Engheta *et al* in their works [17,18].

In conclusion, we have shown how a dielectric split ring can squeeze the electromagnetic energy inside a permeability-near-zero material. In this mechanism, the magnetic field can be enhanced drastically even in a long bended gap. If more current sources are added in domain 2, the absolute energy stream density and magnetic field amplitude in the gap can be even larger. Strong magnetic field has many interesting applications in e.g. sensing and some nonlinear problems. In this process, no resonant enhancement is used. In a similar method, strong electric field can be obtained with an open system without a perfect metallic conductor at the boundary.

This work is partially supported by the National Natural Science Foundation (No. 60890070, and No. 60901039) of China, the Swedish Research Council (VR) and

AOARD.

* sailing@kth.se

**Figure captions:**

FIG. 1. (Color online) The schematic configuration for domains 2 and 3 positioned inside domain 1.

FIG. 2. (Color online) Distributions of (a) the normalized electric amplitude and (b) normalized magnetic amplitude. In (a), the field distribution in domain 2 is shown in the inset for clarity. In (a) and (b), the white point at the center of domain 2 represents the line current. In (b), the energy stream density is also shown in arrows. (c) Normalized electric and magnetic amplitudes at the center of the gap, and (d) normalized output power and energy stream density at the center of the gap, are also shown as the width of the gap varies. In (c), the dotted (red) and solid (black) lines are for the electric and magnetic amplitudes, respectively. In (d), the dotted (red) line is for the power generated by the line current, the solid (black) line is for the power emitted into the exterior space, and the solid (blue) line is for the energy stream density.

FIG. 3. (Color online) Distributions of the normalized electric (a) and magnetic (b) field amplitudes when domain 3 is transformed into a spiral with a long narrow gap. In (b), the energy stream density is also shown in arrows.

FIG. 4. (Color online) Distributions of the normalized electric (a) and magnetic (b) amplitudes when an air slit is opened in the middle of the gap of the structure in Fig.

2(a). In (b), the energy stream density is also shown in arrows, and the inset is an enlarged view of the local area around the air slit (illustrated by the white box).

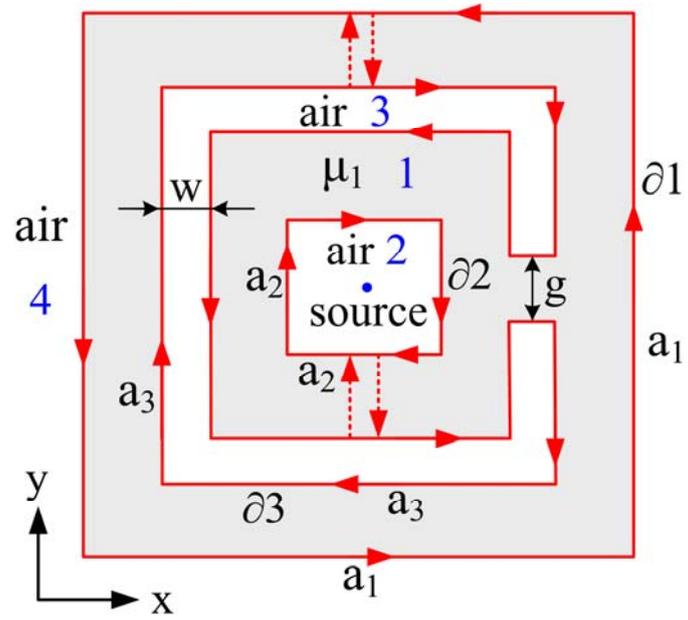

FIG. 1. (Color online) The schematic configuration for domains 2 and 3 positioned inside domain 1.

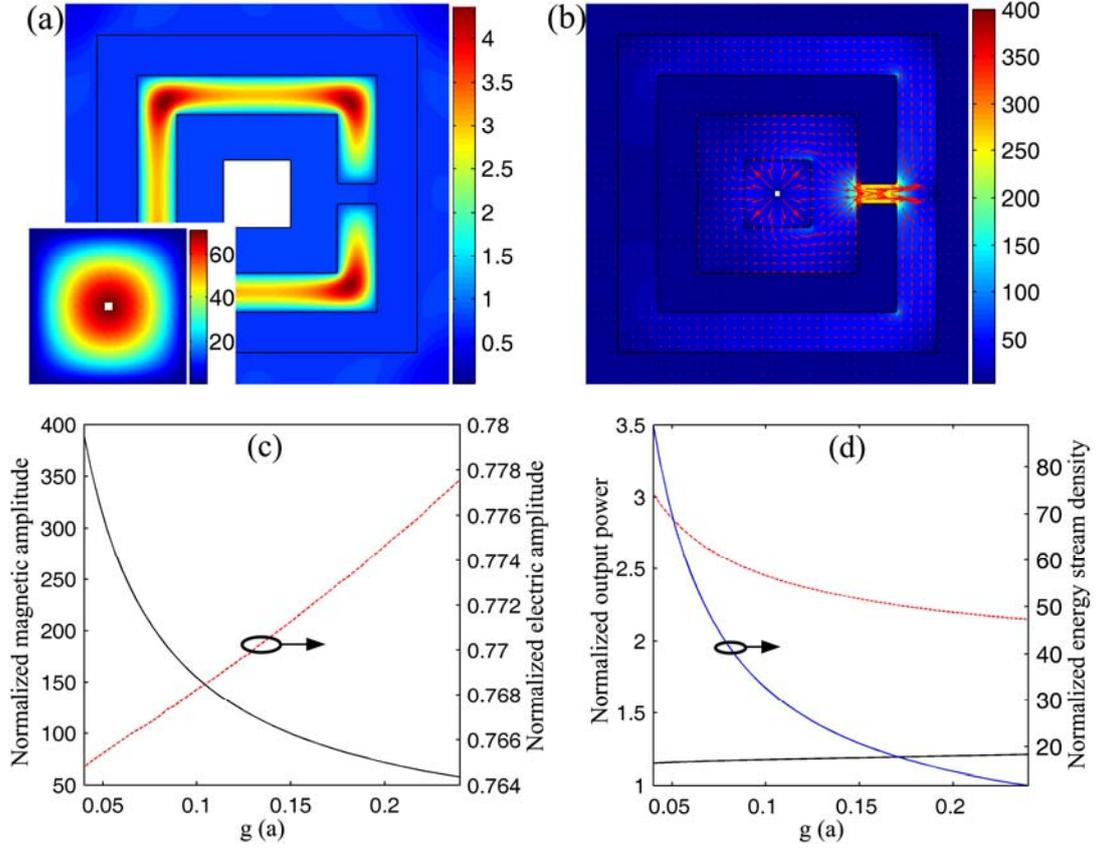

FIG. 2. (Color online) Distributions of (a) the normalized electric amplitude and (b) normalized magnetic amplitude. In (a), the field distribution in domain 2 is shown in the inset for clarity. In (a) and (b), the white point at the center of domain 2 represents the line current. In (b), the energy stream density is also shown in arrows. (c) Normalized electric and magnetic amplitudes at the center of the gap, and (d) normalized output power and energy stream density at the center of the gap, are also shown as the width of the gap varies. In (c), the dotted (red) and solid (black) lines are for the electric and magnetic amplitudes, respectively. In (d), the dotted (red) line is for the power generated by the line current, the solid (black) line is for the power emitted into the exterior space, and the solid (blue) line is for the energy stream density.

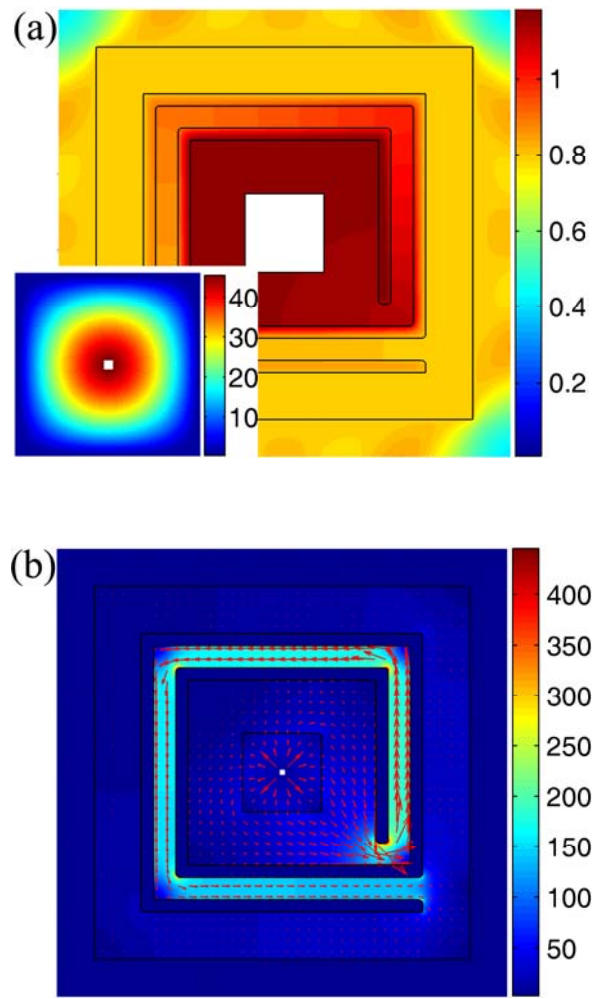

FIG. 3. (Color online) Distributions of the normalized electric (a) and magnetic (b) field amplitudes when domain 3 is transformed into a spiral with a long narrow gap. In (b), the energy stream density is also shown in arrows.

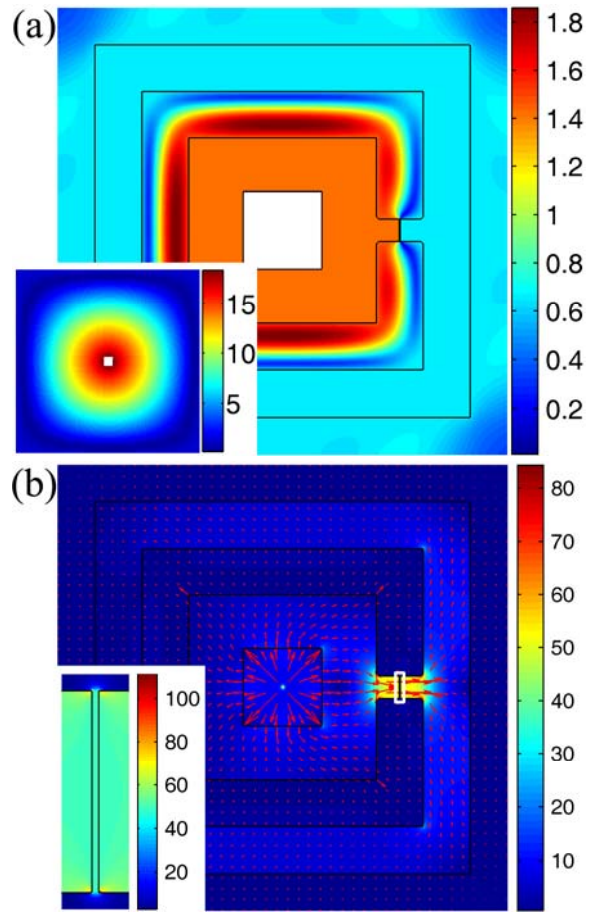

FIG. 4. (Color online) Distributions of the normalized electric (a) and magnetic (b) amplitudes when an air slit is opened in the middle of the gap of the structure in Fig. 2(a). In (b), the energy stream density is also shown in arrows, and the inset is an enlarged view of the local area around the air slit (illustrated by the white box).